# General Solution to Dissipative Waves Considering Attenuation in Time and Space Domains

Peng SHI(时鹏)[1]

[1] Logging Technology Research Institute, China National Logging Corporation, No.50 Zhangba five road, Xi'an, 710077, P. R. China

Corresponding author: sp198911@outlook.com;

*Abstract*

The study points out that the traditional solutions to wave equation of dissipative wave and motion equation of block for a multi-degree-of-freedom mass spring damper system are the possible solutions, which are not necessarily objective and conflict each other. The disturbance in discrete system like crystals vibration can be expressed in differential form. A new general solution to dissipative wave equation is proposed with the general Fourier transform. The solution reveals that the attenuation of the disturbance can simultaneously occur in time and space domains. Then the general solution is used in case studies to analyze the properties of dissipative waves. It is concluded that the properties of waves formulated with the same equation can be different because of the difference of attenuation mechanism.

# 1. Introduction

Wave is an important form of substance movement, which exists widely in nature. According to physical quantities of disturbance, waves are divided into mechanical waves [1], electromagnetic waves [2] and gravitational waves [3], etc. When waves propagate in medium, they transfer the energy and information of disturbance source from one part of medium to another [4, 5]. Since the propagation properties of waves are determined by the properties of the medium and the wave field is affected by the shape and distribution of media, the wave field includes the physical and geometric properties of media [6, 7]. Therefore, by receiving the wave signal propagating in media, the information of disturbance source and the properties of media can be obtained. As the carriers of energy and information, waves have a wide range of applications in many fields including communication, remote controls, medical equipment, machinery manufacturing and earth sciences for information transmission [8, 9], damage detection [6, 7], energy harvesting [10, 11], geophysical exploration [12, 13] and so on.

The study of the general solution to wave equation is an important way to understand the property of wave propagation in a specific medium, especially for linear media, where wave propagation satisfies the principle of superposition. For the wave propagation in a linear medium, its properties can be obtained from the solution of harmonic wave. By viewing the disturbance source as the superposition of harmonic waves with different frequencies, the theoretical solution of the wave caused by arbitrary disturbance in a linear medium can be obtained. When the general solution of harmonic wave and the analytical solution of wave propagation in the linear medium caused by arbitrary disturbances are derived in traditional way, the angular frequency of harmonic waves is always assumed to be a real number [14-18]. It is obtained from the solution that the energy of the disturbance for any medium is only dissipated in space domain. Since the wave equation describes the energy transformation in time domain and energy propagation in space domain, the interaction between adjacent substances exists in both time and space domains in the area where the wave passes by. Therefore, the disturbance dissipation may occur in the time and space domains. As far as I am concerned, where the dissipation comes from is not cleared in physics domain. Therefore, it is not straightforward to say whether the traditional solution to the wave propagation in media with dissipation is correct.

The study tries to give a new general solution to the dissipative waves. Then the general solution is further used to analyze the properties of waves in different dissipative media.

## 2. Discrete system

When mechanical waves propagate in media, the vibration of material element is often treated as a block vibration of mass spring system [14]. In order to smoothly introduce the viewpoint of this paper, we start the work from the analysis of multi-degree-of-freedom mass spring damper system.

Assuming that there is a multi-degree-of-freedom mass spring damper system as shown in Figure 1, the equation of motion of the $n$-th block can be expressed as:

$$m\frac{d^2 u_n}{dt^2} = k\left(u_{n+1} - 2u_n + u_{n-1}\right) + c\left(\frac{du_{n+1}}{dt} - 2\frac{du_n}{dt} + \frac{du_{n-1}}{dt}\right), \quad (1)$$

where $m$ is the mass of block, $u_n$ is the displacement of the $n$-th block, $k$ is the stiffness of spring, $t$ is the time and $c$ is the damping coefficient. If we consider that the distance between blocks $l$ is the unit length and the distance is infinitesimal on the macroscopic scale, Equation (1) can be expressed in differential form as:

$$m\frac{\partial^2 u_n}{\partial t^2} = k\frac{\partial^2 u_n}{\partial x^2} + c\frac{\partial^2 u_n}{\partial x^2 \partial t}, \quad (2)$$

Equation (2) is formally identical to the wave equation of a linear Kelvin–Voigt viscoelastic wave propagating along coordinate axis $x$ [19]. This means that the motion equation of blocks in multi-degree-of-freedom mass spring damper system is equivalent to the wave equation of a dissipative wave. If we realize that infinitesimal is a relative concept rather than an absolute quantity, the concept that Equation (2) is the motion equation of blocks in a multi-degree-of-freedom mass spring damper system can be easily accepted. In fact, the above concept is used in solving the vibration equations of crystals in solid state physics [20].

### 3. Solution to wave equation of dissipative medium

In mechanics of vibration, the motion equation of block expressed like Equation (1) is just a differential equation with respect to time. Therefore, the energy of block vibration is considered to decay only in time domain [21]. However, when the wave equation of dissipative wave is solved, the energy of the disturbance for any medium is considered to decay only in space domain by assuming the angular frequency is a real number [14-18]. That is, for the same equation of motion the processes of dissipation are different in vibration mechanics and wave mechanics. This means that only one of these two assumptions is true or neither is true. At present, the process of dissipation has not been discussed in depth and which process the dissipation exists in is unclear. In mathematics the solution to a dissipative wave equation is not unique and different solutions can be derived from different dissipation mechanisms. Therefore, it is hard to say whether the traditional solution is correct and universal. By considering the propagation of block vibration, the solution to the motion equation of blocks in a multi-degree-of-freedom mass spring damper system may be also not correct and universal.

Since the interaction between adjacent substances exists in both time domain and space domain in the area where the disturbance passes, it is reasonable to assume that the attenuation of vibration occurs in both time and space domains. The general solution can be expressed with the double general inverse Fourier transform as:

$$u(x,t) = -\frac{1}{4\pi^2} \int_{-\lambda-i\infty}^{-\lambda+i\infty} X(p) e^{px} dp \int_{\zeta-i\infty}^{\zeta+i\infty} e^{qt} dq , \qquad (3)$$

where, $p = -\lambda + si$ and $q = -(\zeta + \omega i)$ are complexes, respectively, $s$ is the real wave vector, $\lambda$ and $\zeta$ are the attenuation coefficients related to the propagation and vibration, respectively. In this case, the solution to a wave equation should contain attenuations in vibration propagation and material element vibration and there is no ideal harmonic wave in dissipative media and the amplitude of element vibration decreases with time when the material elements of a dissipative medium vibrate in single frequency. The attenuation coefficients are determined by the attenuation mechanism of the medium to the wave. Mathematically speaking, the value of $\lambda$ and $\zeta$ is uncertain, but their value is determined for a specific material. If the vibration attenuation in unit time is equal to the attenuation in the distance that the wave propagates in unit time. As a result, the following relationship holds,

$$\frac{\zeta}{\lambda} = \frac{\omega}{s} = v, \tag{4}$$

where $v$ is the velocity of wave. When $\zeta=0$, Equation (3) degenerates to the traditional solution of wave with dissipation. When $\zeta=\lambda=0$, the double general inverse Fourier transform in Equation (3) degenerates into the double inverse Fourier transform and Equation (3) degenerates into the solution to non-dissipative waves.

## 4. Case study

Replacing the mass of block $m$, the displacement of the $n$-th block $u_n$, the stiffness of spring $k$ and the damping coefficient $c$ with the density $\rho$, the displacement at a point $u$, the stiffness $K$ and viscosity coefficient $\eta$, correspondingly, the displacement equation of motion for a linear Kelvin–Voigt viscoelastic wave propagating along coordinate axis $x$ is formulated as [19]:

$$\rho \frac{\partial^2 u}{\partial t^2} - K \frac{\partial^2 u}{\partial x^2} - \eta \frac{\partial^3 u}{\partial x^2 \partial t} = 0, \tag{5}$$

Substituting Equation (3) into Equation (5), the dispersion relation of the Kelvin–Voigt

viscoelastic wave is obtained:

$$(\zeta + \omega i)^2 - \frac{K}{\rho}(\lambda - si)^2 + \frac{\eta}{\rho}(\lambda - si)^2 (\zeta + \omega i) = 0. \tag{6}$$

By solving Equation (6) under different dissipation mechanisms, the dispersion relation of the Kelvin–Voigt viscoelastic wave with different dissipation mechanisms can be obtained.

Below we compare the dispersion relations of the Kelvin–Voigt viscoelastic wave when dissipation occurs only in space domain, only in time domain and in both time and space domains. Figure 2 shows the properties of Kelvin–Voigt viscoelastic waves with attenuations only in space domain and in both time and space domains, respectively. It is seen that although the two attenuation mechanisms are different, $\lambda/s$ and velocity have similar trends along the frequency. In the low frequency range ($\omega \ll K/\eta$), $\lambda/s$ and velocity are basically the same for the two different attenuation mechanisms. In high frequency limit ($\omega \gg K/\eta$), $\lambda/s$ tends to $\sqrt{1/3}$ when the attenuations are equivalent in time and space domains ($\zeta/\lambda = v$) and to 1 when the attenuation occurs only in space domain ($\zeta/\lambda = 0$). When the frequency is higher than the critical frequency ($\omega_0 = K/\eta$), the velocity of the two waves increases in the same trend and the velocity of the wave whose attenuation are equivalent in time and space domains is lower than that of the wave whose attenuation occurs only in space domain.

Figure 3 shows the properties of the viscoelastic wave in the Kelvin–Voigt viscoelastic medium with attenuation only in time domain. It is seen that the properties of the viscoelastic wave with attenuation in time domain are obviously different from that of viscoelastic waves with the attenuation in time domain and the attenuations equivalent in time and space domains. There are two propagation modes in the Kelvin–Voigt viscoelastic medium with attenuation in time domain, i. e. underdamped mode (black lines in Figure 3) and overdamped mode (red lines in Figure 3). For the underdamped mode, the velocity decreases with the increase of the frequency whereas $\zeta/\omega$ increases with the frequency. For the overdamped mode, the velocity increases with the frequency, but $\zeta/\omega$ decreases with the increase of frequency. When the frequency tends to the critical frequency, the velocity and $\zeta/\omega$ for the two model tends to $v_0/\sqrt{2}$ and 1, respectively. There is no wave with a higher frequency than the critical frequency exists in the medium. This indicates that the medium turns to be a rigid when frequency is higher than the critical frequency.

**5. Conclusion**

This study points out that the solution to the wave equation describing a wave propagating in a dissipative medium is not unique due to the uncertainty of the dissipation process. A general solution to waves propagating in media with dissipation is proposed by assuming that the attenuation of disturbance occurs in both propagation and vibration progresses. The general solution can degenerate into the traditional solution which considers the attenuation of disturbance only occurs in propagation and describe the motion of blocks in a discrete system. It is revealed that due to the difference of attenuation mechanism the properties of waves formulated with the same equation may be dramatically different and the general solution to the wave equation can be determined only when the disturbance dissipation process is known. The study also shows that infinitesimal is a relative concept rather than an absolute quantity and the disturbance in discrete system can be expressed in differential form.

**Acknowledgments**

This work was supported by the scientific research and technology development project of China National Petroleum Corporation (2020B-3713).

**Statements and Declarations**

There is no competing Interest.

**Figures**

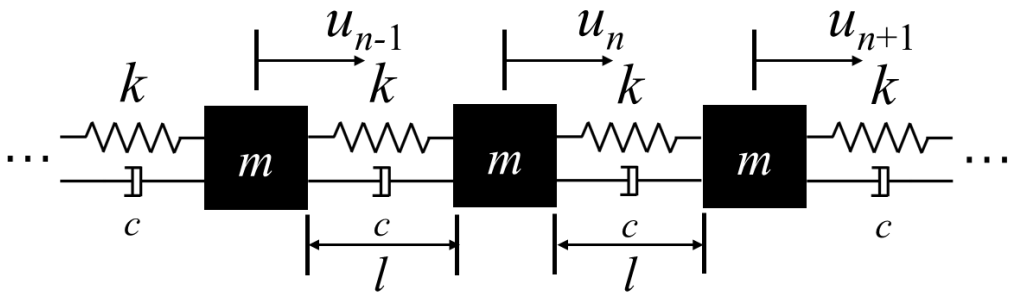

Figure 1. Sketch of multi-degree-of-freedom mass spring damper system.

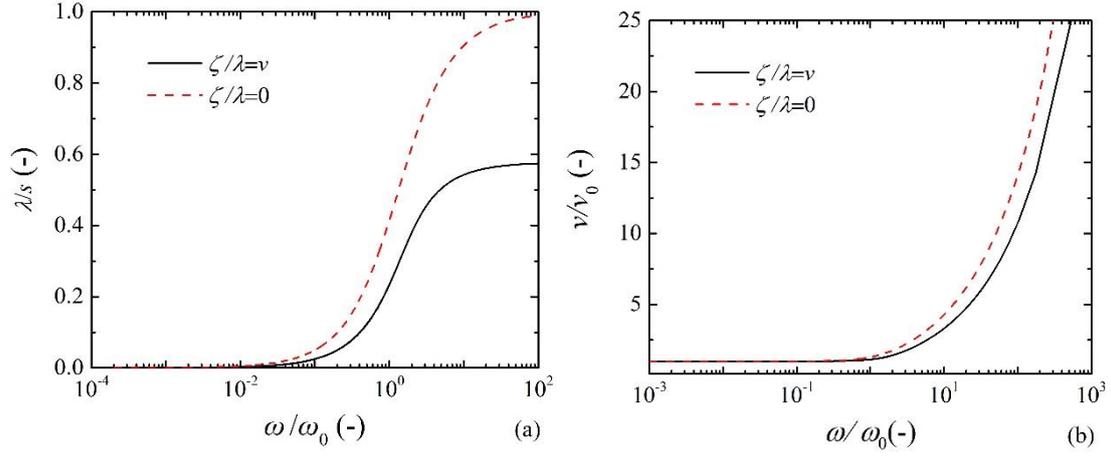

Figure 2. Properties of viscoelastic waves with different attenuation mechanisms. Influence of angular frequency on $\lambda/s$ (a) and velocity (b). $v_0 = \sqrt{K/\rho}$ is the velocity at low frequency limit, $\omega_0 = K/\eta$ is the critical frequency. $\zeta/\lambda = 0$ and $\zeta/\lambda = v$ correspond to the attenuation only in space domain and the attenuations equivalent in time and space domains, respectively.

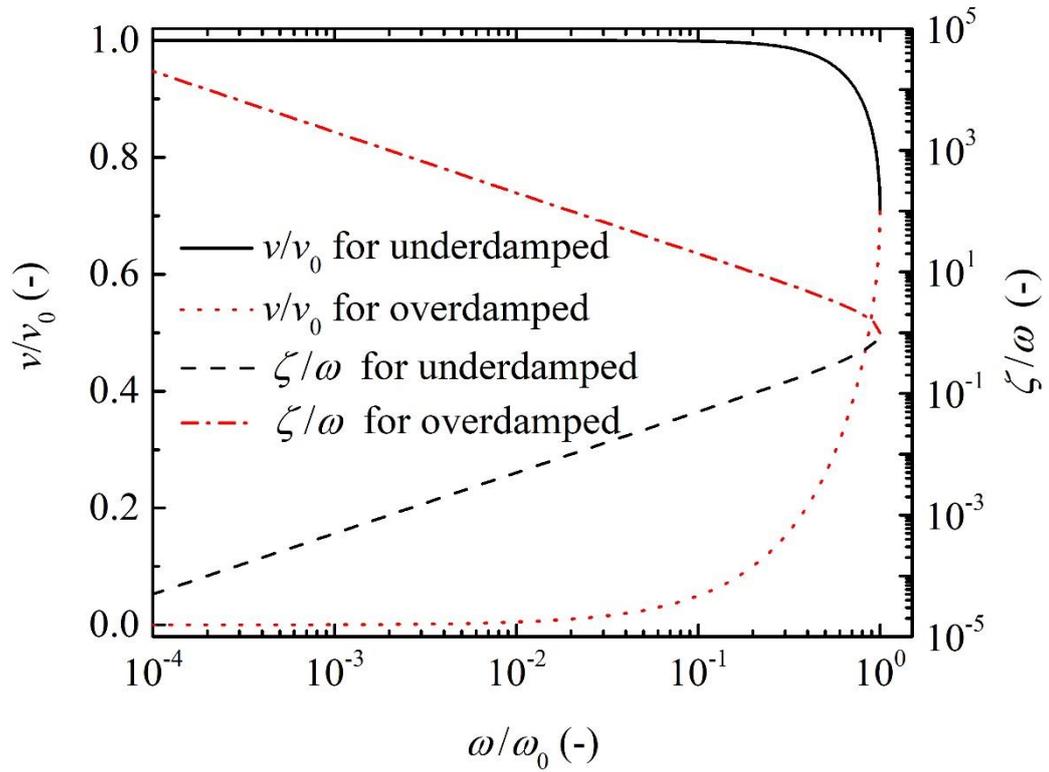

Figure 3. Properties of viscoelastic waves with attenuation in time domain. The black solid and dashed lines represent the influence of frequency on the velocity and $\zeta/\omega$ of underdamped mode, respectively. The red dotted and dash-dotted lines represent the influence of frequency on the velocity and $\zeta/\omega$ of overdamped mode, respectively.